\documentclass[titlepage,12pt]{article}
\usepackage{amssymb,psfig,epsfig,pslatex}

\usepackage{cite}

\makeatletter
\def\@sect#1#2#3#4#5#6[#7]#8{\ifnum #2>\c@secnumdepth
  \def\@svsec{}\else
  \refstepcounter{#1}\edef\@svsec{\csname the#1\endcsname.\hskip0.5em}\fi
  \@tempskipa #5\relax
  \ifdim \@tempskipa>\z@
    \begingroup
      #6\relax
      \@hangfrom{\hskip #3\relax\@svsec}{\interlinepenalty \@M #8\par}%
    \endgroup
    \csname #1mark\endcsname{#7}\addcontentsline
      {toc}{#1}{\ifnum #2>\c@secnumdepth \else
        \protect\numberline{\csname the#1\endcsname}\fi #7}%
  \else
    \def\@svsechd{#6\hskip #3\@svsec #8\csname #1mark\endcsname
      {#7}\addcontentsline{toc}{#1}{\ifnum #2>\c@secnumdepth \else
        \protect\numberline{\csname the#1\endcsname}\fi #7}}%
  \fi \@xsect{#5}}

\newcommand{\one}{1\!\mbox{l}}

\begin{document}

  \begin{flushright}
    CERN-PH-TH/2004-206 \\
  \end{flushright}
\vspace{0.01cm}

\begin{center}
{\LARGE Investigation of Top quark spin correlations\\ at hadron colliders} \\
\vspace{1.2cm}
{\bf W. Bernreuther$^A$, A. Brandenburg$^B$, Z. G. Si$^C$
 and P. Uwer$^D$}
\par\vspace{0.8cm}
{$^A$ Institut f. Theoretische Physik, RWTH Aachen, 
52056 Aachen, Germany\\
$^B$ DESY-Theorie, 22603 Hamburg, Germany\\
$^C$ Department of Physics, Shandong University, Jinan Shandong
    250100\\ P. R. China\\
$^D$Department of Physics, Theory Division, CERN, CH-1211 Geneva 23 \\ Switzerland}
\end{center}
\par\vspace{1cm}

\begin{center}{\bf Abstract} \end{center}
{ We report on our  results
about hadronic 
$t\bar t$ production at NLO QCD including  $t, \bar t$ 
spin effects, especially on $t\bar t$ spin correlations.}

\section{Introduction}

Top quarks, once they are produced in sufficiently large numbers,
are a sensitive probe of the fundamental interactions at high
energies. On the theoretical side this requires precise 
predictions, especially within the
Standard Model (SM). As far as $t \bar t$ production at the Tevatron and LHC is
concerned, spin-averaged (differential) cross section
have been computed   at next-to-leading order (NLO) QCD\cite{nse1,bkns1}
including resummations\cite{Laenen:1993xr,Bonciani:1998vc}.
Observables involving the spin of the 
top quark  can also be calculated
perturbatively, especially within QCD. It is expected that such
quantities  will play an important role in exploring the
interactions that are involved in top quark production and decay.
Within the SM, the QCD-induced correlations between $t$ and $\bar{t}$
spins are large and can be studied at both the Tevatron and the LHC,
for instance  by
means of double differential angular distributions of $t\bar{t}$
decay products. Results for these distributions at 
NLO QCD\cite{Bernreuther:2000yn,Bernreuther:2001bx,Bernreuther:2001rq,Bernreuther:2004npb}
are reviewed below.

\section{Theoretical Framework}

We consider the following processes at hadron colliders
\begin{equation}\label{reaction}
pp/p\bar{p}\rightarrow t\bar{t}+X\rightarrow \left\{
\begin{array}{lcc} l^+l^{'-}&+&X \\
l^{\pm}j_{t(\bar{t})}&+&X\\
j_t j_{\bar{t}}&+&X
\end{array} \right.
\end{equation}
where $l\,\, (l')=e, \mu, \tau$, and $j_t\,\, (j_{\bar{t}})$ denotes the 
jet originating 
from non-leptonic $t\,\,(\bar{t})$ decay. 
At NLO  QCD  the following parton reactions contribute 
to the above processes:
\begin{eqnarray}
&gg, q{\bar q} \ {\buildrel
 t{\bar t}\over \longrightarrow} \  b {\bar b} + 4 f,& \nonumber \\
&gg, q{\bar q} \  {\buildrel
 t{\bar t}\over \longrightarrow} \  b {\bar b} + 4f  + g, &\nonumber \\
&g + q ({\bar q})\  {\buildrel
 t{\bar t}\over \longrightarrow}\   b {\bar b} + 4f  + q ({\bar q}) ,&
\end{eqnarray}
where $f=q,\ell,\nu_{\ell}$.
The calculation of these cross sections at NLO
QCD simplifies 
in the leading pole
approximation (LPA), which is justified because
 $\Gamma_t/m_t < 1\%$.
Within the LPA, the radiative corrections can be classified into
factorizable and non-factorizable contributions. At NLO 
these non-factorizable corrections do not contribute to the double 
differential distributions\cite{Bernreuther:2004npb}, which we will
discuss below.  Therefore we do not consider them here.
Considering only factorizable corrections in the
on-shell approximation 
the squared matrix element $|{\cal M}|^2$ of the respective parton reaction
is  of the form
\begin{equation}\label{eqm}
|{\cal M}|^2 \propto {\rm Tr}\;[\rho R{\bar{\rho}}]
 = \rho_{\alpha'\alpha} R_{\alpha\alpha',\beta\beta'} \;
\bar{\rho}_{\beta\beta'}.
\end{equation} 
Here $R$ denotes the  density matrix that describes
 the production of on-shell
$t\bar t$ pairs in a specific spin configuration, and
$\rho,{\bar{\rho}}$ are the density matrices describing the decay
of polarized $t$ and $\bar t$ quarks, respectively, into specific final states.
The subscripts in  (\ref{eqm}) denote the  $t$, $\bar t$ spin
indices. The spin-averaged production density matrices yield the
NLO cross sections for $t\bar t$ being produced by $q\bar q$, $g g,$
$gq$, and $g \bar q$ fusion\cite{nse1,bkns1}.

To obtain a full NLO QCD analysis of (\ref{reaction}), we must consider
also the NLO QCD corrections to the matrix element of the main SM decay modes
of the (anti)top quark in a given spin state, i.e. the semileptonic modes
$t \to  b \ell^+ \nu_{\ell},$ $b \ell^+ \nu_{\ell} g$  $ (\ell=e,\mu,\tau)$,
and the non-leptonic decays $t \to  b q {\bar q}',$ $
 b q {\bar q}' g$ where  $q {\bar q}'= u {\bar d}, c {\bar s}$ for the
dominant channels. For the computation of the double angular
distributions (\ref{disteq}), the matrix elements of the 2-particle
inclusive parton reactions
$i \ {\buildrel
 t{\bar t}\over \longrightarrow} \ a + b + X $ are required.
Here $a,b$ denote a lepton or a jet. In the LPA this involves the
1-particle inclusive $t$ decay density matrix $2\rho^{t\to a}_{\alpha'\alpha}
= {\Gamma^{(1)}}(\one +{{\kappa}_a}\,{\bf{\tau}} \cdot
{\hat{\bf{q}}_1})_{\alpha'\alpha},$ where ${\hat{\bf{q}}_1}$ is the
direction of flight in the $t$ rest frame and $\Gamma^{(1)}$ is the
partial width of the respective decay channel. An analogous  formula
holds for $\bar t$ decay.
The factor $\kappa_a$  is the $t$ spin analysing power
of particle/jet $a$. 
Its value  is crucial for the
experimental determination of top spin effects, in particular
of $t\bar{t}$ spin correlations. For the standard $V-A$ charged-current
 interactions these coefficients are known to order $\alpha_s$
for semileptonic  \cite{Czarnecki:1991pe} and non-leptonic
\cite{Brandenburg:2002xr} modes.
The charged lepton is  a perfect analyser
of the top quark spin, which is due to the fact that
$\kappa_{\ell}=1-0.015\alpha_s \, .$
In the case of  hadronic top quark
decays, the spin analysing power of jets can be defined, for example
\begin{eqnarray} \label{kappa}
\kappa_b=-0.408\times(1-0.340\alpha_s)=-0.393 \, ,\\
\kappa_j=+0.510\times(1-0.654\alpha_s)=+0.474 \, .
\end{eqnarray}
Here $\kappa_b$ is the analysing power of the $b$ jet
and  $\kappa_j$ refers to  the least energetic non-$b$-quark
jet defined by  the Durham algorithm. Obviously the spin analysing power
is decreased if one uses the hadronic final states to analyse the spins of $t$
and/or $\bar t$. However, this is
(over)compensated by the gain
in statistics and by the  efficiency with which the $t$ (${\bar t}$)
rest frames can be reconstructed.

With the above building blocks, we can discuss the 
following double angular 
distributions\footnote{QCD-generated  absorptive parts in the
parton scattering amplitudes induce a small $t$ and $\bar t$
polarization, which to order $\alpha_s^3$ is normal
to the $q{\bar q}, gg \to t{\bar t}$ scattering planes 
\cite{Bernreuther:1995cx,Dharmaratna:xd}.}
\begin{equation}\label{disteq}
\frac{1}{\sigma}\frac{d\sigma}
{d\cos\theta_1 d\cos\theta_2}=\frac{1}{4}\Big(1-C \cos\theta_1\,\cos\theta_2\Big),
\end{equation}
where $C$ is a measure of
 the $t\bar{t}$ correlations;
 $\theta_{1}\,\, (\theta_{2}) $ is the angle between the 
direction of flight of particle/jet $a_1\,\, (a_2)$
in the $t\,\,(\bar{t})$ rest frame  with respect to
reference directions  $\hat{\bf a}$ ($\hat{\bf b}$), 
which will be specified below.  For the factorizable corrections the
exact formula $C = \kappa_{a_1}\kappa_{a_2} D$ holds\cite{Bernreuther:2001rq}.
Here $D$  is the $t\bar{t}$ double spin asymmetry
\begin{equation}
\label{d}
D=\frac{N(\uparrow\uparrow)+N(\downarrow\downarrow)-N(\uparrow\downarrow)
-N(\downarrow\uparrow)}{N(\uparrow\uparrow)
+N(\downarrow\downarrow)+N(\uparrow\downarrow)+N(\downarrow\uparrow)},
\end{equation}
where $N(\uparrow\uparrow)$ denotes the number of $t\bar{t}$ pairs
with $t$ ($\bar{t}$) spin parallel to the reference axis $\hat{\bf a}$ 
($\hat{\bf b}$),  etc.
Thus $\hat{\bf a}$ and $\hat{\bf b}$  
can be identified with the
quantization axes of the $t$ and $\bar{t}$ spins, respectively, 
and $D$ directly 
reflects the strength of the correlation between the 
$t$ and $\bar{t}$ spins for the chosen axes. 

For $t\bar t$ production at the Tevatron it is well known that 
the so-called off-diagonal
basis\cite{Mahlon:1997uc}, which is defined by the requirement that 
$\hat{\sigma}(\uparrow\downarrow)$ $=\hat{\sigma}(\downarrow\uparrow)=0$
for the process $q\bar{q}\to t\bar{t}$ at tree level, yields 
a large coefficient $D$. It has been shown in \cite{Bernreuther:2001rq}
that the beam basis, where $\hat{\bf a}$ and $\hat{\bf b}$ are
identified with the hadronic beam axis, is practically as good as the
off-diagonal basis.  A further possibility is the helicity basis,
which is a good choice for the  LHC.

\section{Predictions for the Tevatron and the LHC}

We now discuss the  spin correlation coefficients
$C$  of the distributions (\ref{disteq}).
It should be noted that beyond LO QCD, 
it is important to construct infrared and collinear safe observables at
parton level. In the case at
hand it boils down to the question of the frame  in which  the
reference directions $\hat{\bf a}$ and $\hat{\bf b}$ are to be
defined. It has been  shown  that, 
apart from the $t$
and $\bar t$ rest frames,  the $t\bar t$ zero
momentum frame (ZMF) is the appropriate frame for defining
collinear safe spin-momentum observables. 
The off-diagonal, beam, and helicity bases  are defined  in
the $t \bar t$ ZMF.  Details can be found in  Ref.\cite{Bernreuther:2004npb}.

In Table~\ref{tab:Tevlhc} we list our predictions 
for  $C$ in 
 (\ref{disteq}) at the Tevatron and LHC.
The results are obtained using the CTEQ6L (LO) 
and CTEQ6.1M (NLO) parton distribution functions
(PDF) \cite{Pumplin:2002vw}. 
Numbers are given for the dilepton (L-L), lepton$+$jet (L-J) and
all-hadronic (J-J)
decay modes of the $t\bar{t}$ pair, in the latter two cases 
the least energetic non-$b$-quark jet (defined by the Durham cluster
algorithm) was used as spin analyser. 
One notices that  for the  Tevatron  the spin correlations are largest in the
beam and off-diagonal bases, and the QCD corrections reduce the LO results
for the coefficients $C$ by about 10\% to 30\%. For the LHC 
the QCD corrections are small ($<10\%$).
These results are obtained with 
$\mu\equiv \mu_R=\mu_F=m_t=175$ GeV. 
At the Tevatron, a variation of the scale $\mu$ 
between $m_t/2$ and $2m_t$  changes the  results at $\mu=m_t$
by $\sim \pm$ (5--10)\%, while at the LHC  the change of $C_{\rm hel}$ is 
less than a per cent.
\begin{table}[htbp!]
\caption{\it LO and NLO results for the spin correlation coefficients
${\rm C}$  of the distributions
(\ref{disteq}) for the Tevatron at $\sqrt{s}=1.96$ TeV and 
for LHC at $\sqrt{s}=14$ TeV.
The PDF  CTEQ6L (LO) and CTEQ6.1M
(NLO) were used, and
$\mu_F=\mu_R=m_t=175GeV$.}\label{tab:Tevlhc}
\begin{center}
\renewcommand{\arraystretch}{1.2}
\begin{tabular}{|ccccc|} \hline
          &    &  L--L  &L--J  & J--J  \\ \hline
\multicolumn{5}{|c|}{Tevatron} \\ \hline
${\rm C}_{\rm hel}$ &LO  & $-0.471$    & $-0.240$ & $-0.123$ \\
          &NLO & $-0.352$    & $-0.168$ &  $-0.080$ \\ \hline
${\rm C}_{\rm beam}$ &LO &  $\phantom{-}0.928$
&  $\phantom{-}0.474$ &
$\phantom{-}0.242$ \\
           &NLO&  $\phantom{-}0.777$    &  $\phantom{-}0.370$
&  $\phantom{-}0.176$ \\ \hline
${\rm C}_{\rm off}$   &LO&  $\phantom{-}0.937$
&  $\phantom{-}0.478$
&  $\phantom{-}0.244$ \\
           &NLO&  $\phantom{-}0.782$    &   $\phantom{-}0.372$
&  $\phantom{-}0.177$ \\ \hline
\multicolumn{5}{|c|}{LHC} \\ \hline
${\rm C}_{\rm hel}$ &LO  &   $\phantom{-}0.319$
&  $\phantom{-}0.163$
&  $\phantom{-}0.083$ \\
          &NLO &   $\phantom{-}0.326$   &
$\phantom{-}0.158$
&   $\phantom{-}0.076$ \\ \hline
\end{tabular}
\end{center}
\end{table}

In Table~\ref{tab:PDF} we compare the NLO results for the spin correlation
coefficients evaluated for the CTEQ6.1M, MRST2003 \cite{Martin:2003sk}
and GRV\cite{Gluck:1998xa} PDFs. It is easy to see that the results 
with  the recent CTEQ6.1M and MRST2003 PDF agree
at the per cent level (this is not the case for previous versions
of the CTEQ and MRST PDF), while the GRV98 PDF 
gives significantly different results at the Tevatron.
This shows that the spin correlations are very sensitive to the
relative quark and gluon contents of the proton\cite{Bernreuther:2001rq}. 
Future measurements  of (\ref{disteq})
may offer the possibility
to further constrain the quark and gluon contents of the proton. 
\begin{table}
\caption{\it Spin correlation coefficients at NLO for different PDFs for
the Tevatron (upper part) and the LHC (lower part) for dilepton final
states.}\label{tab:PDF}
\begin{center}
\renewcommand{\arraystretch}{1.2}
\begin{tabular}{|cccc|} \hline
\multicolumn{4}{|c|}{Tevatron} \\ \hline
              &  CTEQ6.1M  &MRST2003       &GRV98\\ \hline
${\rm C}_{\rm hel}$ & $-0.352$    &  $-0.352$   & $-0.313$  \\  \hline
${\rm C}_{\rm beam}$  &  $\phantom{-}0.777$
&   $\phantom{-}0.777$
& $\phantom{-}0.732$   \\ \hline
${\rm C}_{\rm off}$  &  $\phantom{-}0.782$
& $\phantom{-}0.782$   & $\phantom{-}0.736$ \\ \hline
\multicolumn{4}{|c|}{LHC} \\ \hline
${\rm C}_{\rm hel}$ &    $\phantom{-}0.326$   &
$\phantom{-}0.327$   &  $\phantom{-}0.339$ \\ \hline
\end{tabular}
\end{center}
\end{table}

Before closing this section, we summarize how 
an experimental measurement of the distributions 
(\ref{disteq}) that matches our predictions
should proceed: 
1)  Reconstruct the top and antitop 4-momenta in the laboratory frame
  ($=$ c.m. frame of the colliding hadrons).
2)  Perform a rotation-free boost from the laboratory frame to the 
  $t\bar{t}$~ZMF. Compute ${\bf \hat a}$ and ${\bf \hat b}$ in that frame.
3) Perform rotation-free boosts from the $t\bar{t}$~ZMF
  to the $t$ and $\bar t$ quark rest frames.  Compute the direction
  ${\bf \hat q}_1$ $({\bf \hat q}_2)$ of the $t$ $(\bar t)$ decay
  product $a_1$ $(a_2)$
in the $t$ $(\bar t)$ rest frame. 
  Finally, compute
  $\cos\theta_1={\bf \hat a}\cdot{\bf \hat q}_1,\  
  \cos\theta_2={\bf \hat b}\cdot{\bf \hat q}_2$.
Here one should notice that in this prescription 
the $t$ and $\bar{t}$ rest frames are obtained by first boosting
into the $t\bar{t}$~ZMF. If this step is left out, and the 
$t$ and $\bar{t}$ rest frames are constructed 
by directly boosting from the lab frame, a Wigner rotation has to
be taken into account.

\section{Conclusion}
\label{concl}
We have computed at NLO QCD
 the $t\bar{t}$ spin correlations in hadronic top production,
which are large effects
within the SM. Our present results are obtained without imposing 
kinematic cuts. Such cuts will in general distort the distributions, 
i.e. ${C}$ will in general depend on the angles $\theta_1$ 
and $\theta_2$.
One strategy is to correct for 
these distortions by Monte Carlo methods before extracting the
spin correlation coefficient and comparing it with theoretical predictions.
A future aim is to directly include the cuts
in an NLO event generator to be constructed with our NLO results
for all relevant  $2\to 6$ and $2\to 7$ processes.

\section{Acknowledgement}
This work was  supported by the
Deutsche Forschungsgemeinschaft, SFB-TR9 (W.B.),  by a Heisenberg
fellowship (A.B.), and by the Chinese National Science Foundation and CSC
(Z.G.S.).

\end{document}